\documentclass[prl,twocolumn,superscriptaddress,floatfix,amsmath,nofootinbib,amssymb]{revtex4}

\usepackage{graphicx}
\usepackage{dcolumn}
\usepackage{mathrsfs}
\usepackage{bm}
\usepackage{color,xspace}
\usepackage[small,compact,raggedright]{titlesec}
\usepackage{epstopdf}

\newcommand{\tr}{\operatorname{tr}}

 \newcommand{\beq}{\begin{equation}}
\newcommand{\eeq}{\end{equation}}
\newcommand{\bea}{\begin{eqnarray}}
\newcommand{\eea}{\end{eqnarray}}

\newcommand{\bra}{\langle}
\newcommand{\ket}{\rangle}

\newcommand{\bb}{\mathbf}

\begin{document}

\title{Persistence of Tripartite Nonlocality for Non-inertial Observers}
 \author{Alexander Smith and Robert B. Mann}
\affiliation{Department of Physics, University of Waterloo, Waterloo, Ontario Canada N2L 3G1}

\begin{abstract}
We consider the behaviour of bipartite and tripartite non-locality between fermionic entangled states shared by observers,  one of whom uniformly accelerates.  We find that while fermionic entanglement persists for arbitrarily large acceleration, the Bell/CHSH inequalities cannot be violated for sufficiently large but finite acceleration.
However the Svetlichny inequality, which is a measure of genuine tripartite non-locality, can be violated for any
finite value of the acceleration.
\end{abstract}


\pacs{04.70.Dy, 03.65.Ta, 04.62.+v, 42.50.Dv}


\maketitle

 The relationship between entanglement and non-locality is at the root of the foundations of quantum mechanics.
Bell-type inequalities \cite{Bell}, such as the Clauser-Horner-Shimony-Holt
(CHSH) inequality \cite{Clauser},  place an upper bound on the correlations compatible with  classical local realistic (or hidden variable) theories.  All   pure entangled states of two qubits violate the CHSH inequality, with the amount of violation increasing  with the degree of bipartite entanglement \cite{Gisin}.  For pure tripartite-entangled states the Svetlichny inequality \cite{Svetlichny} functions in a similar way --  its violation is a sufficient condition for the confirmation of genuine three-qubit nonlocal correlations.  It has recently been shown that a class of generalized GHZ (GGHZ) states do not violate the Svetlichny inequality if their 3-tangle is less than 1/2, whereas another class of states known as maximal slice (MS) states violate the Svetlichny inequality with the amount of violation increasing with the degree of tripartite entanglement \cite{ghose}.

In this paper we investigate the behaviour of nonlocal correlations of bipartite and tripartite entangled states shared between observers in a situation where one observer moves with uniform acceleration.  Previous work has shown that  both bipartite \cite{alsing,fuentes} and tripartite \cite{Hwang,WangJing} entanglement are degraded as a function of increasing acceleration of a given detector.  For fermionic states this degradation persists to a finite value
\cite{alsing,WangJing} in the large acceleration limit in both cases.   We find that whereas bipartite  nonlocal correlations vanish for finite values of the acceleration, tripartite nonlocal correlations persist for any finite acceleration for both GGHZ and MS three-qubit states provided the respective control parameters are  appropriately chosen, vanishing only in the infinite acceleration limit.  This demonstrates that tripartite entanglement and its associated nonlocal correlations are more robust to relativistic effects.

We consider a  flat spacetime in which there is a uniformly accelerating observer, Rob. The two other observers, Alice and Charlie,  remain inertial. Each observes the same state of a  fermionic field (described by a Grassmann scalar, the simplest case preserving the essential Dirac characteristics \cite{EduIvy}). However while both Charlie and Alice use a basis of Minkowski modes (plane waves in the massless case) for the description of their respective parts of the field state, Rob uses the so-called Rindler modes \cite{Dragan}. Uniformly accelerated detectors couple to these modes. The  change of basis between Rindler and Minkowski modes (given by the so-called Bogoliubov coefficients  \cite{Dragan,basis})  mixes Minkowski creation and annihilation operators. Consequently the Minkowski and Rindler vacua are not the same, giving rise to the well-known Unruh effect \cite{Unruh}.

Rindler coordinates $(\eta,\xi,y,z)$ describe a family of  observers with uniform acceleration $a$ and divide Minkowksi spacetime (with coordinates   $(t,x,y,z)$)  into 4 regions separated by a light cone centered on the origin and labelled clockwise as regions F, I, P, and II;  
Rightward accelerating observers are  located in region I (where $t=\xi \sinh(a\eta/c)$ and $x=\xi \cosh(a\eta/c)$) and are causally disconnected from their analogous counterparts in region II.   Monochromatic solutions of the field equation relevant to the accelerating detector in Rindler coordinates are called Rindler modes.  The annihilation operator in region I for the Rindler mode of a given particle species of  frequency $\omega$, and spin-$\sigma$ component is denoted $c_{\omega,\sigma,I}$, with its antiparticle counterpart denoted $d_{\omega,\sigma,I}$, and its creation operators obtained from the dagger of these.  Region II has   corresponding pairs of annihilation/creation operators.  
From the accelerated observer's viewpoint the Hilbert space factorizes as ${\cal H}_I \otimes {\cal H}_{II}$. Since Rob (by definition) is confined to  region I, we must trace out the part of the state outside of this region (ie we trace over all states in region $ {\cal H}_{II}$), since he cannot observe it.  Analogous considerations apply to observers in region II, generally referred to as anti-Rob \cite{alsing}.

An inertial observer would express Rob's vacuum mode $\left|0\right\rangle_R$, and first excitation of his vacuum mode $\left|1\right\rangle_R$ as  \cite{alsing}
\bea
  \left|0\right\rangle_R &=& \cos r \left|0\right\rangle_I \left|0\right\rangle_{II} + \sin r \left|1\right\rangle_I \left|1\right\rangle_{II}, \label{R0} \\
  \left|1\right\rangle_R &=& \left|1\right\rangle_I \left|0\right\rangle_{II}.
  \label{R1}
\eea
where $\left|0\right\rangle_{I}$ and $\left|1\right\rangle_{I}$ are respectively the Minkowski vacuum and first excitation of the Minkowski vacuum in region I; $\left|0\right\rangle_{II}$ and $\left|1\right\rangle_{II}$ are defined similarly over region II.  For simplicity we consider all orthogonal modes as unexcited. The parameter
$r$ is 
\beq
\cos r = \frac{1}{\sqrt{1+\exp(-2\pi \Omega)}},
\eeq
where $\Omega= \frac{\omega c}{a}$ and $a$ is Rob's acceleration, $c$ is the speed of light and $\omega$ is the central frequency of the fermion wave packet. Note that $r=0$ corresponds to an inertial observer and $r=\pi/4$ corresponds to an observer in the infinite acceleration limit. 
 
It is possible to construct a different set of modes that have the property that they are
linear combinations of purely positive-frequency Minkowski modes.   Referred to as Unruh modes, they
are given by \cite{Dragan}
\begin{eqnarray}
\tilde{c}^\dagger_{\omega,\sigma,R} &=& \cos r {c}^\dagger_{\omega,\sigma,I} - \sin r 
{d}^\dagger_{\omega,-\sigma,II} \nonumber\\
 \tilde{c}^\dagger_{\omega,\sigma,L} &=& \cos r {c}^\dagger_{\omega,\sigma,II}  - \sin r
{d}^\dagger_{\omega,-\sigma,I} \label{umodes}
\end{eqnarray}
where $\tan r  = e^{-\pi\Omega}$.  The subscripts L and R denote  left and  right  modes, and are related to each other by an exchange of regions I and II.  In terms of creation and annihilation operators, this means that (\ref{umodes}) can be rewritten as a linear combination of only Minkowski creation operators, implying that the Minkowski and Unruh vacua are the same. 

Hence we can write the Minkowski vacuum $\left| 0 \right\rangle_M$ as
$
\left| 0 \right\rangle_M = \bigotimes_{\omega}  \left| 0 \right\rangle_{\omega, U}
$
where $\tilde{c}^\dagger_{\omega,\sigma,R}$ and $\tilde{c}^\dagger_{\omega,\sigma,L}$ each annihilate
$\left| 0 \right\rangle_{\omega, U}$ for every $\sigma$.  We can therefore restrict our considerations to 
a particular frequency $\omega$.  An arbitrary Unruh excitation is 
\begin{equation}
\left| 1 \right\rangle_{\omega, U} = 
\left(q_R \tilde{c}^\dagger_{\omega,\sigma,R} + q_L \tilde{c}^\dagger_{\omega,\sigma,L} \right)\left| 0 \right\rangle_{\omega, U} 
\end{equation}
with $ |q_R|^2 + |q_L|^2 = 1$.  We shall set $q_L=0$, (the so-called single mode approximation \cite{singlemode}), and henceforth drop the redundant subscripts $\omega$, $U$, and $\sigma$.

 The well-known CHSH inequality \cite{Clauser}
\beq
\left|C(\bb{a},\bb{b}) + C(\bb{a}',\bb{b}) + C(\bb{a},\bb{b}') - C(\bb{a}',\bb{b}')\right| \leq 2
\eeq
quantifies  bipartite non-locality,   where $\bb{a}$ and $\bb{a}'$ are a set of unit vectors belonging to one observer and $\bb{b}$ and $\bb{b}'$ are unit vectors belonging to another observer who are measuring the spin along these directions. The correlation functions $C(\bb{a},\bb{b})$ are given by
\beq
C(\bb{a},\bb{b}) = \left(\frac{2}{\hbar}\right)^2 \left\langle \psi \right| \bb{a}\cdot S_1 \otimes \bb{b} \cdot S_2 \left|\psi \right\rangle.
\label{correlation1}
\eeq
where $S_1$ and $S_2$ are the spin operators acting on modes 1 and 2 respectively. Let us consider the singlet state
\beq
\left|\psi\right\rangle = \frac{1}{\sqrt{2}} \left( \left|10\right\rangle - \left|01\right\rangle \right)
\label{state}
\eeq
 invariant under the unitary operators which generate rotations of coordinate systems. 
 This means that $C(\bb{a},\bb{b})$ is a function of $\cos \theta_{ab}:=\bb{a}\cdot\bb{b}$ only, and hence there is no loss of generality in assuming that $\bb{a}$ points along the $z$ axis and that $\bb{b}$ lies in the $x$-$z$ plane. Then equation (\ref{correlation1}) becomes
\beq
C(\bb{a},\bb{b}) = \left\langle \psi \right| \sigma_{1z} \otimes \left(\sigma_{2z} \cos \theta_{ab} + \sigma_{2x} \sin \theta_{ab} \right) \left|\psi \right\rangle.
\label{correlation2}
\eeq

An accelerating observer making measurements in a sufficiently small interval about $\eta=0$ will use the same
vectors as in eq. (\ref{correlation1}) , but   would see qubit 2   described by equations (\ref{R0}) and (\ref{R1}). 
Constructing the  reduced density operator $\rho_{A,I} = \tr_{II}[\left|\psi\right\rangle\left\langle \psi \right|]$
by tracing out region II  we obtain from (\ref{correlation2})
\beq
C(\bb{a},\bb{b}) = - \frac{1}{2} \left( 1 + \cos 2 r\right) \cos \theta_{ab}.
\eeq
Let us now restrict our attention to the special case in which (i) the vectors $\bb{a}$, $\bb{a}'$, $\bb{b}$, $\bb{b}'$ are co-planar; (ii) $\bb{a}$ and $\bb{b}$ are parallel; and (iii) $\theta_{ab'}=\theta_{a'b}=\gamma$. Then the CHSH inequality will be satisfied provided
\beq
   \cos^2r  \left|\sin^2 \gamma+ \cos \gamma \right| \leq 1
\label{bell}
\eeq
yielding a specific threshold acceleration $a_t = \frac{2\pi\omega c}{\ln 4}$  above which   the CHSH inequality cannot be violated, despite the persistence of fermionic entanglement for $a>a_t$ \cite{alsing}.

Turning next to tripartite nonlocality,  we consider the generalized GHZ (GGHZ) state $\left|\psi_g\right\rangle$ and the maximal slice (MS) state $\left|\psi_s\right\rangle$  
\bea
\left|\psi_g\right\rangle &=& \cos \theta_1 \left|000\right\rangle + \sin \theta_1 \left|111\right\rangle  \label{gghz} \\
\left|\psi_s\right\rangle &=& \frac{1}{\sqrt{2}} \left\{ \left|000\right\rangle + \left|11\right\rangle  \left[\cos \theta_3 \left|0\right\rangle +\sin \theta_3 \left|1\right\rangle \right]\right\}  \label{ms}
\eea
where $\theta_1$ and $\theta_3$ are parameters controlling the entanglement of the state. The usual GHZ state is realized by setting $\theta_1=\pi/4$ in   (\ref{gghz}) or by setting $\theta_3=\pi/2$ in   (\ref{ms}).  To avoid difficulties in  distinguishing violations arising from 2-body versus 3-body entanglement \cite{ghose,Collins,cereceda}, we shall consider the Svetlichny inequality, as its violation is a sufficient condition for the confirmation of genuine 3-qubit nonlocal correlations \cite{ghose, Svetlichny}. This inequality emerged from a consideration of a 
 model of nonlocal-local-realism \cite{Svetlichny},  where two of the qubits are nonlocally correlated but are locally correlated to the third, implying  an entanglement between the three qubits that cannot be decomposed into a case of familiar two body entanglement.

Suppose we have a system that can be partitioned into three subsystems, respectively detectable by Alice, Charlie and Bob. 
Alice performs   the measurements $A=\hat{\bb{a}}\cdot \vec{\sigma}_1$ or $A'=\hat{\bb{a}}'\cdot \vec{\sigma}_1$  on system 1,
Charlie performs $C=\hat{\bb{c}}\cdot \vec{\sigma}_2$ or $C'=\hat{\bb{c}}'\cdot \vec{\sigma}_2$ on system 2 and Bob performs
measurements  $B=\hat{\bb{b}}\cdot \vec{\sigma}_3$ or $B'=\hat{\bb{b}}'\cdot \vec{\sigma}_3$   on system 3, where $\vec{\sigma}_i$ are projection operators that can be expressed in terms of the Pauli operators and $\hat{\bb{a}}$, $\hat{\bb{a}}'$, $\hat{\bb{b}}$, $\hat{\bb{b}}'$, $\hat{\bb{c}}$, $\hat{\bb{c}}'$ are unit vectors corresponding to the direction of measurement. For a theory  consistent with Svetlichny's nonlocal-local-realism 
\beq
S(\Psi):= \left| \left\langle \psi \right|S\left|\psi\right\rangle\right| \leq 4,
\eeq
where $S$ is the Svetlichny operator defined as
\beq\label{Svet}
S = A(CK + C'K') + A'(CK - C'K').
\eeq
where $K=B+B'$ and $K'=B-B'$. 
 
To maximize the expectation value of $S$ for the the GGHZ and MS states \cite{ghose}, we set $\bb{a}=(\sin \theta_a \cos \phi_a, \sin \theta_a \sin \phi_a, \cos \theta_a)$ and similarly for $\hat{\bb{a}}'$, $\hat{\bb{b}}$, $\hat{\bb{b}}'$, $\hat{\bb{c}}$ and $\hat{\bb{c}}'$ .Defining   unit vectors $\hat{\bb{d}}$ and $\hat{\bb{d}}'$ such that $\hat{\bb{c}} + \hat{\bb{c}}' = 2 \hat{\bb{d}} \cos \theta$ and $\hat{\bb{c}} - \hat{\bb{c}}' = 2 \hat{\bb{d}} \sin \theta$, with $D = \hat{\bb{d}} \cdot \vec{\sigma}_2$ and $D' = \hat{\bb{d}}' \cdot \vec{\sigma}_2$, the expectation value of (\ref{Svet}) is
\bea
S(\psi) &=& 2\left|\cos \theta \left(\bra ADB \ket - \bra A'DB' \ket \right) \right.           \nonumber \\
 &&\quad  \left.
+ \sin \theta \left(\bra AD'B' \ket + \bra A'D'B \ket  \right)\right|           \nonumber \\
        &\leq& 2 \left|\left(\bra ADB \ket^2+ \bra AD'B' \ket^2\right)^{1/2} \right.           \nonumber \\
 &&\quad  \left.
        + \left(\bra A'D'B \ket^2+ \bra A'DB' \ket^2\right)^{1/2}\right|  \label{S}
\eea
using the identity $x \cos \theta + y \sin \theta \leq (x^2 +y^2)^{1/2}$.

Replacing Bob by the uniformly accelerating Rob, who observes qubit 3 in a sufficiently small interval about $\eta=0$, we construct the density operator, $\rho=\left|\psi\right\rangle \left\langle \psi\right|$, from (\ref{R0}) and (\ref{R1}) and use this to evaluate  (\ref{S}), yielding
\begin{widetext}
\bea
S(\psi_g) &\leq& 2 \left\{ \left( 2\cos^2 \theta_1 \cos^2r - 1 \right)^2  \left(\cos^2 \theta_d+\cos^2 \theta_{d'}\right)\cos^2 \theta_a\right. 
 + \left.\sin^2 2\theta_1 \cos^2 r \left(\sin^2 \theta_d+\sin^2 \theta_{d'}\right)\right\}^{1/2} \sin^2 \theta_a \nonumber \\
 &+& 2\left\{ \left( 2\cos^2 \theta_1 \cos^2r - 1 \right)^2 \left(\cos^2 \theta_d+\cos^2 \theta_{d'}\right) \cos^2 \theta_{a'}\right. 
 + \left.\sin^2 2\theta_1 \cos^2 r \left(\sin^2 \theta_d+\sin^2 \theta_{d'}\right) \sin^2 \theta_{a'}\right\}^{1/2} 
\eea
\end{widetext}
for  the GGHZ state (\ref{gghz}).

\begin{figure}
	\centering
		\includegraphics[scale=0.4]{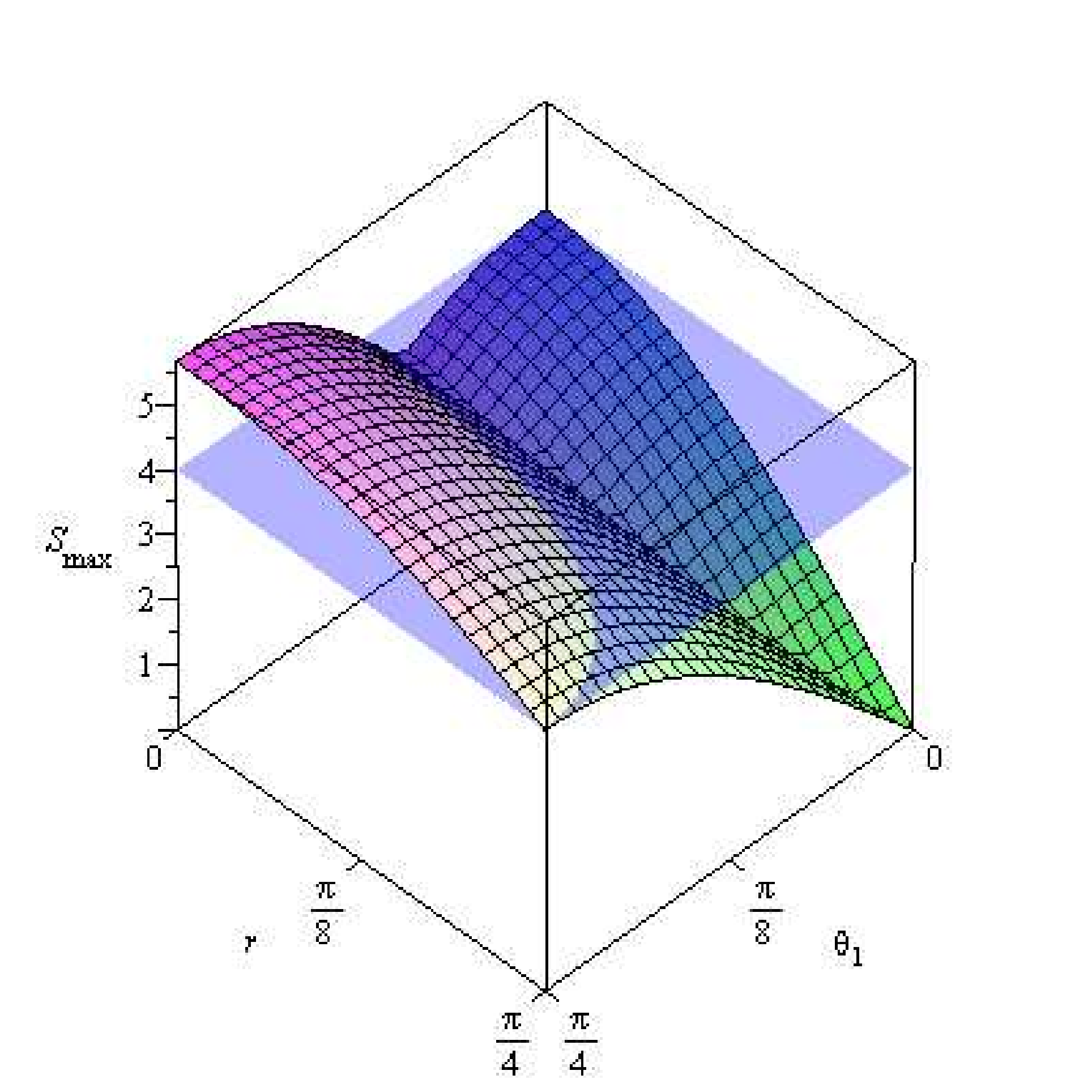}
	\caption{The maximal value of the left hand side of the Svetlichny inequality  plotted against Rob's acceleration $r$ and the control parameter $\theta_1$ for the GGHZ state. As $\theta_1$ increases so does the $\pi$-tangle of the state.}
\label{fig:svetGGHZ}
\end{figure}

Maximizing with respect to $\theta_a$ and $\theta_{a'}$ and 
 noting that the maxima of $\cos^2 \theta_d+\cos^2 \theta_{d'}$ and $\sin^2 \theta_d+\sin^2 \theta_{d'}$ are respectively 1 and 2 yields
 \beq
S(\psi_g) \leq
\left\{
     \begin{array}{lr}
       4\left( 2\cos^2 \theta_1 \cos^2r - 1 \right) & : C_3 \geq C_4\\
       4\sqrt{2}\sin 2\theta_1 \cos r  &  : C_3 < C_4
     \end{array}
   \right. \label{sgghz} .
\eeq
where
\beq
C_3 = \left( 2\cos^2 \theta_1 \cos^2r - 1 \right)^2 \quad
C_4 = \sin^2 2\theta_1 \cos^2 r 
\eeq
As $r\to 0$ Rob, the accelerated observer, becomes the inertial observer Bob and eq. (\ref{sgghz}) agrees with the results of \cite{ghose}. We plot in figure \ref{fig:svetGGHZ}  the value of $S$ when the inequality in (\ref{sgghz}) is saturated as a function of the control parameter $\theta_1$ and Rob's acceleration $r$.
A straightforward (but lengthy) calculation of the $\pi$-tangle  (a measure of tripartite entanglement for mixed states \cite{WangJing,Ou})
indicates that $\theta_1$ monotonically increases as a function of the $\pi$-tangle.  Hence we find for all finite values of the acceleration parameter that nonlocality increases as a function of increasing entanglement for the GGHZ state beyond a certain threshold value of the $\pi$-tangle.  Note that more entanglement is required to violate the Svetlichny inequality as $a$ increases; it is not possible to violate the inequality in the infinite acceleration limit. 

For the MS state its reduced symmetry between only  qubits 1 and 2  implies that there are two cases to consider for one accelerating observer: (i) when Rob is measuring either qubit 1 or 2 and (ii) when Rob is measuring qubit 3. 

\begin{figure}
	\centering
		\includegraphics[scale=0.4]{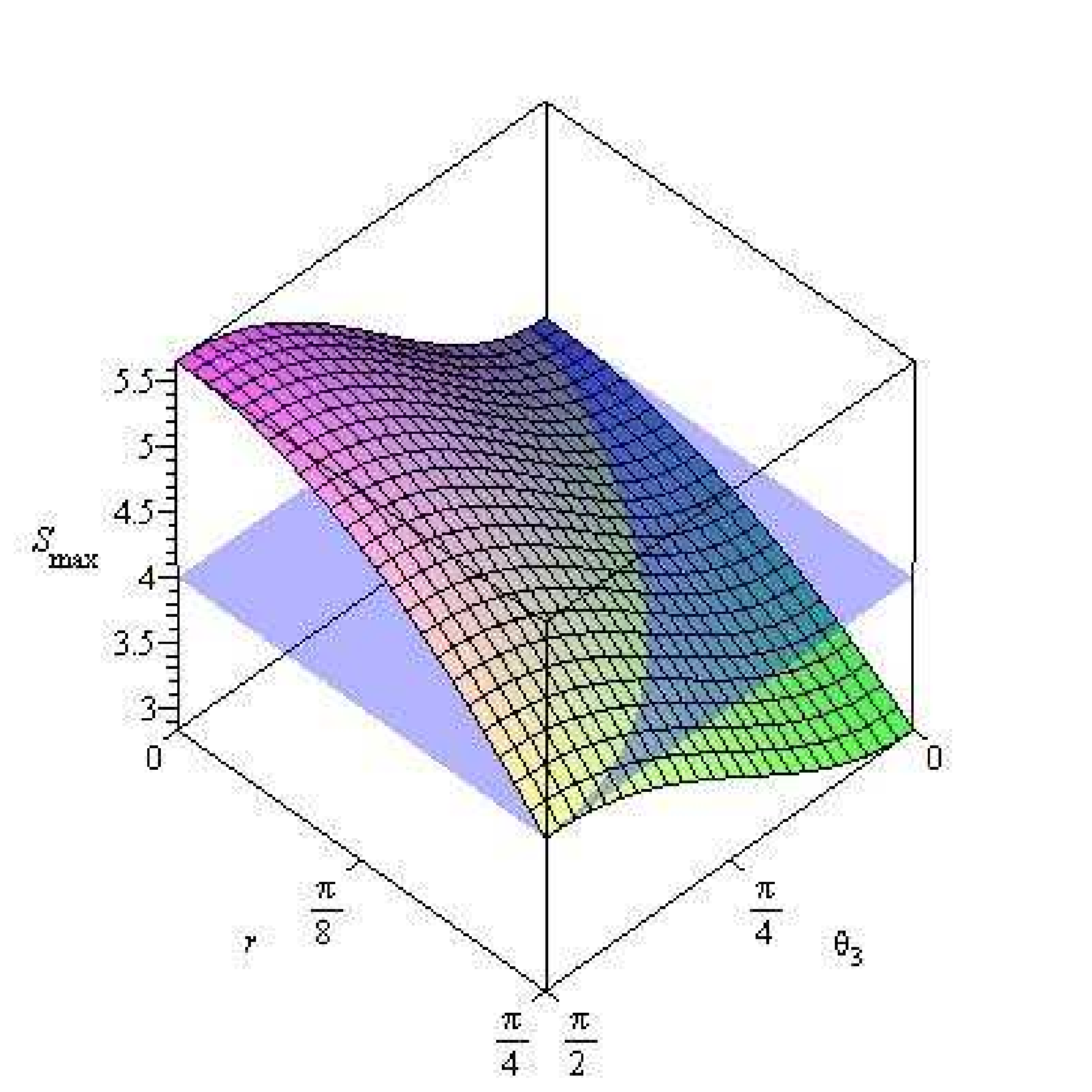}
	\caption{The maximal value of $S(\psi_M)$  plotted as a function of the control parameter $\theta_3$ and Rob's acceleration $r$,  with Rob  measuring qubit 2.}
	\label{fig:svetMS1}
\end{figure}

For case (i) we construct the density operator  by substituting equations (\ref{R0}) and (\ref{R1}) into the second qubit of the MS state in equation (\ref{ms}). Tracing out region II  to obtain the density operator $\rho_{AIC}$ and using this to evaluate 
eq. (\ref{S}) gives
\beq
S(\psi_s) \leq 4 \cos r \left\{\cos^2 \theta_3 + 2\sin^2 \theta_3 \right\}^{1/2} \label{sms1}
\eeq
where  $S(\psi_s)$ has a maximum at $\phi_d-\phi_{d'}=\theta_d=\theta_{d'}=\pi/2$ \cite{ghose}.
The maximal value of $S(\psi_s)$ from eq. (\ref{sms1}) is plotted in figure \ref{fig:svetMS1}.
A calculation indicates that $\theta_3$ monotonically increases as a function of the associated $\pi$-tangle. In this case we see that for any finite value of the acceleration  there exists a value of the  $\pi$-tangle (or $\theta_3$) for which the  maximal value of $S(\psi_s)$ is larger than 4 and the
Svetlichny inequality is violated.   

For case (ii)  the analogous result is  
\beq
S(\psi_s) \leq 4 \left\{\cos^2 \theta_3  \cos^2 2r  +2\sin^2 \theta_3 \cos^2 r\right\}^{1/2}
\eeq
and it is straightforward to show that the  maximal value of $S(\psi_s)$ is also larger than 4 for all finite values of $r$ provided $\theta_3$ is appropriately chosen.
 
For increasing uniform acceleration, violation of both the bipartite Bell-CHSH and tripartite Svetlichny inequalities
becomes increasingly difficult, despite the survival of fermionic entanglement in the large $a$ limit.  For
$a>a_t$ all bipartite nonlocality vanishes, whereas tripartite nonlocality persists for all finite values of $a$. 
This suggests  a fundamental difference between tripartite and bipartite entanglement in relativistic settings.
 \bigskip

R.B.M was supported in part by
the Natural Sciences and Engineering Research Council of Canada.  

\textbf{Note added} As this work was being completed  reference \cite{Educoll} appeared,
in which similar results were obtained for the Bell/CHSH inequalities.

\end{document}